# Female Fertility and Longevity

Joshua Mitteldorf
Dept of Ecology and Evolutionary Biology
University of Arizona
Tucson, AZ 85920
(215) 248-1010 voice;   (530) 678-5456 fax
*josh@mathforum.org*

## Abstract

Does bearing children shorten a woman's life expectancy?  Several demographic studies, historic and current, have found no such effect.  But the Caerphilly cohort study is far the most prominent and frequently-cited, and it answers in the affirmative.  Why has this study found an effect that others fail to see?  Their analysis is based on Poisson regression, a statistical technique that is accurate only if the underlying data are Poisson distributed.  But the distribution of the number of children born to women in the Caerphilly database departs strongly from Poisson at the high end.  This makes the result overly sensitive to a handful of women with 15 children or more who lived before 1700.  When these 5 women are removed from a database of more than 2,900, the Poisson regression no longer shows a significant result.  Bi-linear regression relating life span to fertility and date of birth results in a *positive* coefficient for fertility.

# Female Fertility and Longevity

Joshua Mitteldorf

## 1. Introduction

Evolutionary theory offers powerful reasons to believe that an individual's fertility should be inversely related to life span. This 'cost of reproduction' is predicted by all the pleiotropic theories which have dominated the field of life history evolution for more than fifty years. According to these theories, aging has evolved despite its negative effect on individual fitness because of tradeoffs between fertility and longevity. There is powerful selection for alleles that offer fertility enhancement, and some of these have side-effects that permit long-term degradation of the soma.

In Williams's (1957) original formulation of Antagonistic Pleiotropy, the tradeoff was genetic; it is the *potential* for reproduction that has exacted a longevity cost. Alternatively, the Disposable Soma (DS) theory of Kirkwood (1977) is based on a direct metabolic tradeoff between fertility and longevity. Hence the DS theory in particular predicts there should be a steep and direct cost of reproduction. Within the DS theory, individuals that bear fewer offspring are predicted to live longer than those who bear many offspring, and this remains true regardless of whether curtailment is endemic or contingent.

Confirmation of this prediction has been elusive. Some studies showing positive correlations between fertility and longevity, while most find no significant association. In a large-scale, contemporary study of Norwegians, Grundy (2008) found a positive correlation between fertility and longevity. Perls (1997) found a strong positive relationship with late childbearing in a study of Boston centenarians.

By the logic of the DS theory, the effects of modern contraception should not affect the expectation of a negative correlation. Nevertheless, several researchers have sought to avoid any confounding effect of technological civilization by looking to historical databases. The DS theory ought to apply equally to males and to females, but association between male fertility and longevity is also consistently positive (Palmore 1982; Kaplan 1988; Davey Smith, Frankel et al. 1997), so that some researchers seeking confirmation of the theory have limited their analyses to historic data on human females.

Korpelainen (2000) examined historic data from rural Finns and European aristocrats in the 18[th] and 19[th] centuries. She found positive correlations between number of offspring and life span in both men and women, with the (expected) negative effect only for a small sample of women over 80. Lycett et al (2000) studied a North German historical population in the same time frame, and found, again, a "reverse cost" of reproduction, particularly strong among the lower socio-economic classes. Le Bourg et al (Le Bourg, Thon et al. 1993) (also (Muller, Chiou et al. 2002)) studied French-Canadian populations in this same time frame and found no evidence for a cost of reproduction. (These studies and other demographic data are reviewed by Le Bourg (2001). McArdle *et al* (2006) studied an Amish population from the 18[th] through the 20[th] centuries, and found a positive relationship between fertility and longevity. After removing the inverse correlation associated with long-term demographic trends, a positive correlation for men remained, and the correlation for women was insignificant.

The one large-scale human study which claims to discern a negative correlation was published by Kirkwood himself together with J. G. Westendorp (1998). They analyzed women from the Caerphilly database of British aristocracy going back 800 years. Because this publication appeared prominently (in *Nature*), and because it corroborated the theory, it has been widely cited as authority. There are more citations of the Caerphilly study than all the other demographic papers combined (as listed in ISI Web of Science as of October, 2008). For this reason, revisiting the Caerphilly methodology seems worthwhile.

## 2 The Peerage Database

Westendorp and Kirkwood take for their source a published database with family histories of British aristocracy. Their choice of an upper-class sample was made to lower the noise in the data from accidental deaths related to poverty or hardship; historical rather than present data was preferred because the relationship they sought concerned inborn fecundity, and could be masked by contraceptive practices that have become common in the twentieth century. Their analysis was limited to women because the reproductive physiology of males is less directly related to the actual number of children sired. The authors further limit analysis to married women for whom there is complete data, who were born before 1875, and who died after menopause (taken alternatively as 50, 55 or 60 years of age); they sought to correlate the number of children each woman bore with her age at death.

The message of their result is announced in the title, "Human longevity at the cost of reproductive success." Elsewhere, they couch their conclusions in more guarded language. "The relation between age at death and progeny number in British aristocratic women (and men) is consistent with the hypothesis that the longest lived individuals have reduced fertility compared with the majority of the population. This effect is seen most strongly in women born before 1700, for whom the number of children was larger than for women born between 1700 and 1875." (Westendorp and Kirkwood 1998, p. 745)

The sophisticated statistical methods which these authors bring to bear on the data uncover an inverse relationship between fertility and longevity. But it is also true that a less sophisticated analysis points to the opposite conclusion.

### Alternative analysis with linear regression

Linear regression of age at death vs. number of children produces a correlation coefficient ($r=-0.008\pm0.018$) that is negative but not significant. The clearest signals in the data are that over the centuries, longevity increases ($r=0.247\pm0.018$) while fertility decreases ($r=-0.126\pm0.018$). An approach based on analysis of variance suggests a two-variable regression, age at death vs number of children and year of birth, to segregate this effect. When this is done, the result is a marginally significant *positive* relationship between number of children and age at death ($r=0.025\pm0.019$). Westendorp (Westendorp 2001) has suggested that a more sensitive correction for social trends in fertility may be made if from the number of children each woman has borne is subtracted the average number of children borne by a sliding group of her contemporaries. When this is done, the positive correlation becomes significant (r=0.060±0.018).

(The AP predictions ought to apply equally well to women and to men. The corresponding analysis for men in the database again yields a significant positive correlation between fertility and longevity, r=0.021±0.011).

Fig. 1 displays age at death for women, averaged by the number of their children. There is no clear downward trend in the chart, even without correction for year of birth. (Data for the chart and regression calculations derive from a later edition of the same source (Bloore 2000) used by Westendorp and Kirkwood, a CD ROM listing family records of British Peerage. Included are the 2919 women born before 1906 for whom exact birth dates and death dates are available, and who survived to their 55th birthdays.)

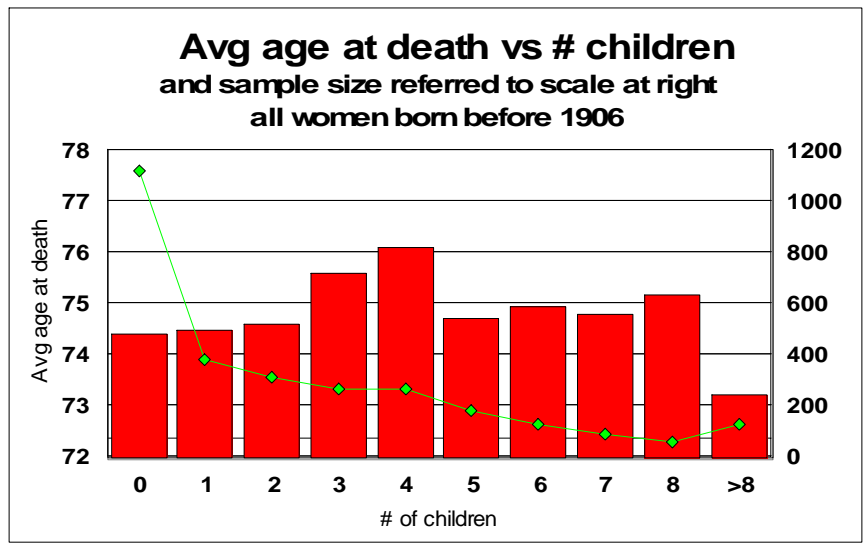

Fig. 1. Age at death for women, averaged by the number of their children. There is no clear downward trend in the chart, even without correction for year of birth. The line indicates sample size, referred to the scale at the right.

### 2.3 Why the disparity in conclusions?

Westendorp's only significant results arise when the sample is limited to women born before 1700. This subsample is relatively small, numbering only 241 out of the full 2919 in the database that meet other inclusion criteria. The group in which they claimed to detect a measurable deficit in fertility comprises the longest-lived subset of those early women, who attained the age of 80 or more. There were 26 such women, and their mean number of progeny was 2.65±0.76, compared with 3.26±0.28 for women who lived 55 to 79 years. (Uncertainties here are computed as standard deviation of the mean.)

But the small sample size does not fully explain the disparity between the present results and Westendorp's: a linear regression limited to women born before 1700 produces an insignificant result (r=−0.14±0.46), and a bar chart limited to these women shows no clear trend (Fig. 2).

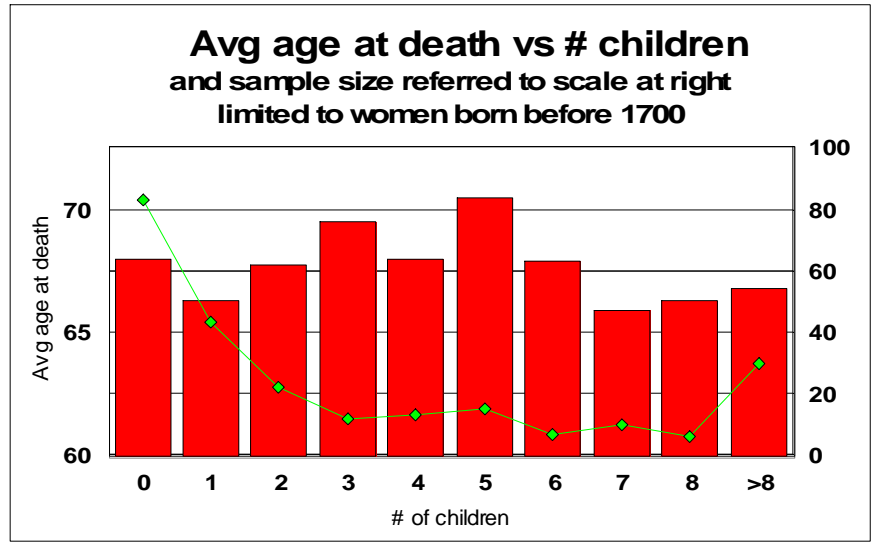

Fig 2. Same as figure 1, limited to women born before 1700. It is in these women that Westendorp and Kirkwood claim to discern a downward trend at the right side of the chart. Note the small sample size, indicated in the scale at the right.

A more important explanation for the disparity in results is that we use linear regression while Westendorp and Kirkwood use a less familiar tool, the Poisson regression. Poisson regression can be used to measure an association between any independent variable that is continuous, and a dependent variable that is Poisson distributed. The Poisson character of the dependent variable is an essential precondition; when the dependent variable is not Poisson distributed, the method can produce anomalous results. A Poisson distribution is controlled by a single parameter, and the core of Poisson regression analysis is to optimize over that parameter in such a way that likelihood of the observed distribution is a maximum.

For the Westendorp analysis, the calculation would proceed thus: Assume that the probability that a given woman will bear n children derives from a Poisson probability P(n,x), where the Poisson parameter x is a linear function of her age at death, x=A*age+B. Try different values for A and B, and calculate a product of probabilities for each woman that she would have borne the number of children that she did in fact bear. Home in on those values of A and B that maximize the probability product. Westendorp's result is that A is barely negative for women in their sample born before 1875, and is significantly negative when the sample is limited to women born before 1700. They report these results after allowance was made for trends over time in both fertility and longevity.

It is possible that a major reason for the disparity between Westendorp's Poisson result and the linear regression analysis is that number of children born to a woman is not a Poisson variable. Fig. 3 shows the actual distribution of children for each woman in the sample, with a mean-adjusted Poisson curve overlaid, where x=2.36 was the actual mean number of children per woman in the sample. Compared to the Poisson curve, there is an excess for n=0 births, and a deficiency for n=1 through 4.

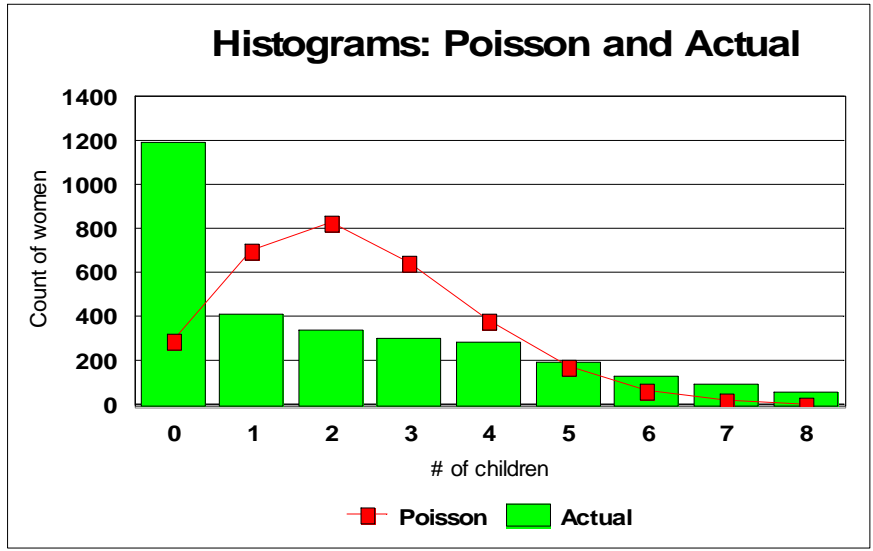

Fig 3. For Poisson regression to be a valid trend indicator, the underlying data must be Poisson distributed. The figure compares the distribution of women in the full sample by number of children with a Poisson distribution of the same mean.

More important to the Poisson regression calculation is the excess of women with large numbers of children. The Poisson probability becomes exponentially small for high n, but there are substantial numbers of women in the sample with more than 10 children. For example, in the sample of 2919, the most prolific woman bore 18 children. The Poisson probability that there should be such a woman is $10^{-7}$. There are 9 women with 15 or more children, while the Poisson function predicts 0.0001 such women.

These observations suggest a reason for the difference between the Poisson and linear results. The Poisson probability that is maximized in Westendorp's regression calculation is inordinately influenced by the women at the very end of the distribution. Even though their numbers are small, the contribution that they make to the aggregate probability is dominant. If these few women should happen to have short lives, that would be sufficient to tilt the Poisson regression result to a negative coefficient. The average age at death of the 9 women with the most children was 67.6, compared to 74.7 for the sample as a whole, a disparity that is well explained by the average birth year of the 9: 1677, compared to 1830 for the whole sample. If these 9 women are removed from the sample of 2919, the Poisson regression no longer generates a negative coefficient. 5 of the 9 women with the most children were born before 1700. The conclusion of Westendorp and Kirkwood depends critically on this handful of women; if they are removed from the analysis, Poisson regression on the remaining subjects produces an insignificant result, even for the pre-1700 data set.

## 4. Laboratory evidence

These results should be viewed in the context of a growing body of animal data, which also fails to support the theoretical notion of a cost of reproduction. In his encyclopedic review of the literature, Finch (1990) reported that there was considerable animal evidence for an immediate mortality cost of reproduction, but little evidence that reproduction affects the rate of aging. Stearns (1992) has appended a survey of animal tradeoff studies to his text on life histories, cataloguing evidence for and against the existence of a tradeoff. Genetic experiments with nematodes suggests that genes affecting the rate of aging are not necessarily linked to fertility (Partridge and Gems 2002); while work with fruitflies has been interpreted as supportive of tradeoffs (Partridge and Gems 2002), though the data on their face demonstrate that breeding for longevity actually enhances fertility (Mitteldorf 2004). Reznick *et al* (2000) review evidence that longevity and fertility in water fleas appear to be independently variable. Most recently, Ricklefs (2007) conducted a large-scale study of 18 mammalian species and 12 avian species in captivity, and found no correlation between fertility and longevity.

## 5. Discussion

The pleiotropic theories posit that tradeoffs are the root cause of aging. In Williams's (1957) original formulation of Antagonistic Pleiotropy, it is the potential for reproduction that is enhanced by pleiotropic alleles, so that the link between actual fertility and longevity may be less absolute. Still we should expect to find a negative correlation in any large, unbiased sample.

But the DS theory in particular predicts unequivocally that fertility should have a clear negative impact on life span, regardless of the origin of variation in fertility. This is a central prediction that may be taken as a fair test of the theory's core. Neither animal experiments nor human demographic data have corroborated the theoretical prediction of a strong cost of fertility.